\documentclass[aps,twocolumn,showpacs,showkeys,amsmath,amssymb,floatfix]{revtex4}
\usepackage{graphicx}
\usepackage{epsfig}
\usepackage{dcolumn}
\usepackage{bm}
\usepackage{amssymb}
\usepackage{dsfont}
\usepackage{amsmath}
\newcommand{\be}{\begin{equation}}
\newcommand{\ee}{\end{equation}}
\newcommand{\bea}{\begin{eqnarray}}
\newcommand{\eea}{\end{eqnarray}}
\newcommand{\hn}{\hat n}
\newcommand{\hatm}{\hat m}
\newcommand{\tC}{{\tilde C}}
\newcommand{\vX}{{\vec X}}

\newcommand{\pro}{\partial}

\newcommand{\D}{{\hat D}}

\newcommand{\ksi}{{\hat \xi}}

\newcommand{\ba}{\begin{array}}
\newcommand{\ea}{\end{array}}

\newcommand{\nn}{\nonumber}

\newcommand{\vH}{{\vec H}}
\newcommand{\vC}{\vec C}
\begin{document}

\title{Weyl symmetric structure of QCD vacuum}
\bigskip
\author{Y. M. Cho}
\affiliation{School of Electrical and Computer Engineering  \\
Ulsan National Institute of Science and Technology, Ulsan 689-798, Korea}
\affiliation{College of Natural Sciences, Department of Physics and Astronomy,
Seoul National University, Seoul 151-747, Korea}
\author{D. G. Pak}
\affiliation{ Institute of Modern Physics, Chinese Academy of Sciences,
 Lanzhou 730000, China }
\affiliation{Lab. of Few Nucleon Systems,
Institute for Nuclear Physics, Ulughbek, 100214,
     Uzbekistan}
 \author{P. M. Zhang}
\affiliation{ Institute of Modern Physics, Chinese Academy of Sciences,
 Lanzhou 730000, China }
\author{L. P. Zou}
\affiliation{ Institute of Modern Physics, Chinese Academy of Sciences,
 Lanzhou 730000, China }


\begin{abstract}
We consider Weyl symmetric structure of the classical vacuum
in quantum chromodynamics. In the framework of formalism
of gauge invariant Abelian projection we show that
classical vacuums can be constructed in terms of
Killing vector fields on the group $SU(3)$. Consequently,
homotopic classes of Killing vector fields
determine the topological structure of the vacuum.
In particular, the second homotopy group $\pi_2(SU(3)/U(1)\times U(1))$
describes all topologically non-equivalent vacuums which are classified
by two topological numbers. Starting with a given Killing vector
field one can construct vacuums forming a Weyl sextet representation.
An interesting feature of $SU(3)$ gauge theory is that it admits
a Weyl symmetric vacuum represented by a linear superposition of
the vacuums from the Weyl vacuum sextet.
A non-trivial manifestation of the Weyl symmetry is demonstrated
on monopole solutions. We construct a family of finite energy monopole solutions
in Yang-Mills-Higgs theory which includes the Weyl monopole sextet.
We conjecture that a similar Weyl symmetric vacuum structure can
be realized at quantum level in quantum chromodynamics.
\end{abstract}
\vspace{0.3cm}
\maketitle

\section{Introduction}

The mechanism of confinement in quantum chromodynamics (QCD)
based on the Meissner effect in dual color superconductor
is very attractive \cite{nambu, mandelstam, polyakov77},
and many features of quark confinement are described in numerous
approaches to low energy QCD in agreement with experimental data.
Despite on this the origin of color confinement remains much less known up to now.
Formally, from the mathematical point of view the color confinement
is manifestation of the fact that color symmetry represents an exact symmetry
of strong interaction. This raises a simple, but fundamental question:
why $SU(3)$ color symmetry in QCD is preserved, whereas
$SU(2)$ gauge symmetry of weak interaction
is spontaneously broken? A possible answer to that question can be
related to features of the groups $SU(2)$ and $SU(3)$,
namely, to Weyl symmetry and its physical
implications in classical and quantum vacuum structures.

In the present paper we study the structure of the classical vacuum and
related issues on monopole solutions in $SU(3)$ QCD.
In a standard approach the classical vacuum configurations are described
by pure gauge potentials classified by the third homotopy group $\pi_3(SU(N))=Z$,
i.e., by the topological Chern-Simons number (see, for ex.,
a review \cite{shuryak} and refs. there in).
Such a non-trivial topological vacuum structure is
manifested through the vacuum tunneling effect realized by means
of instantons \cite{bpst,callan,jackiw}.
Our approach to study of the vacuum structure
is based on the gauge invariant Abelian projection
proposed originally in
\cite{choprd80, choprl81, duan} and developed further
in \cite{periw,fadd1,shab,zhang}.
An essential feature of the formalism of Abelian projection
is that it allows to describe the topological properties of the
vacuum fields in terms of a more simple
geometric object than the gauge potential,
namely, in terms of Killing vector field $\hat m_i$, $i=1,...,N-1$,
on the group $SU(N)$.
In the case of $SU(2)$ gauge theory the classical vacuums
can be classified by the knot number (Hopf number) corresponding to
the third homotopy group $\pi_3(SU(2)/U(1))$
\cite{baal,choplb07}. Due to equivalence
of the homotopy groups $\pi_3(SU(2))\simeq\pi_{2,3}(SU(2)/U(1))$
one has one to one correspondence between topological non-equivalent classes
for the gauge potential and the Killing vector field.
The case of $SU(3)$ gauge theory
reveals a more rich topological content of field configurations.
Even though, the third homotopy groups $\pi_3(SU(N))=Z$ for
$N=2$ and $N=3$ are the same, the second homotopy groups
$\pi_2 (SU(2)/U(1))=Z$ and
$\pi_2 (SU(3)/U(1)\times U(1))=Z\times Z$
describing homotopic classes of
Killing vector fields are essentially different.
This implies important consequences:
{\it (i) topological classical vacuum structure
in $SU(3)$ QCD is determined by two topological numbers;
(ii) topologically non-equivalent vacuums in $SU(3)$ case form
a Weyl sextet of degenerated vacuums and a
non-trivial Weyl symmetric vacuum singlet.}

To find physical implications of the non-trivial topological vacuum
structure we first demonstrate that $SU(3)$
topologically non-equivalent vacuums
form representations of the Weyl symmetry group.
Starting with a given Killing vector field
one can construct a Weyl vacuum sextet representation.
A remarkable feature of $SU(3)$ QCD is that
there exists a Weyl symmetric vacuum.
It is important to stress that such a non-trivial
vacuum does not exist in $SU(2)$ gauge theory.
Another interesting fact is that singular Wu-Yang type
monopole solutions are classified due to Weyl representation theory
as well \cite{goddard,choprl80,choplb82}.
We consider Weyl structure of finite energy monopole solutions in Yang-Mills-Higgs
theory. Introducing a more general one-parameter family of
monopole solutions in the BPS limit we show that different topological
classes of monopoles are separated by infinite energy barrier.

The presence of Weyl symmetric structure
of the classical vacuum and monopole solutions
indicates to possible existence of a similar vacuum structure
with monopole condensation in quantum theory.
There are some indications that two-loop effective potential in QCD
may admit the Weyl sextet of degenerated vacuums \cite{gies,nedelko}.
This may shed light on the origin of color confinement phenomenon in QCD.
We conjecture that to preserve the color symmetry in QCD
against spontaneous symmetry breaking
there must exist a non-trivial Weyl symmetric vacuum in addition to
the Weyl vacuum sextet in the full quantum theory.
In one-loop approximation such a Weyl symmetric vacuum
does exist \cite{flyv,mpla2006}, this provides a possible
stable monopole condensation \cite{prd02,jhep04}.
As another application of our approach based on gauge invariant
Abelian projection we consider a general ansatz for searching
essentially $SU(3)$ instanton and monopole solutions.

The paper is organized as follows.
In Section II we overview briefly Cho-Duan-Ge gauge invariant Abelian projection
in $SU(3)$ gauge theory. We propose an alternative parametrization for Killing
vectors in terms of two complex triplet fields which allows to describe
the geometric origin of the Killing vectors and dual magnetic symmetry.
In Section III we describe the topological vacuum structure in $SU(3)$ QCD
and provide an explicit construction of Weyl representations for
topologically non-equivalent vacuums. A detailed analysis of instanton
solution with a general ansatz including two topological
numbers is presented in Section IV. Section V is devoted to
Weyl symmetric structure of singular and finite energy monopole solutions.
The last section contains conclusions and discussion of quantum vacuum structure
in QCD.

\section{Cho-Duan-Ge gauge invariant Abelian projection}

Let us start with main outlines of
Cho-Duan-Ge gauge invariant Abelian projection in $SU(3)$ QCD
\cite{choprd80, choprl81, duan}.
A principal role in the construction of the Abelian projection belongs to Killing vector fields
$\hatm^a_i,~(a=1,...8,~i=3,8)$ which
describe all mappings from the base space-time to
the homogeneous coset space ${\cal M}^6=SU(3)/U(1)\times U(1)$.

The Abelian decomposition of $SU(3)$ gauge connection
is given by \cite{choprd80, choprl81, duan}
\bea
&& \vec A_\mu =\hat A_\mu + \vec X_\mu \nn \\
&& \hat A_\mu = A^i_{\mu} \hatm_i + \vec C_\mu \nn \\
&& \vec C_\mu^a =-f^{abc} \hatm^{b}_i \pro_\mu \hatm^{c}_i \equiv
 -(\hatm_i \times \pro_\mu \hatm_i)^a,  \label{Adec}
\eea
where $\hat A_\mu$ is a restricted potential, $A_\mu^i$ is Abelian
"photon" gauge potential,
$\vec C_\mu$ is a magnetic potential,
and $\vec X_\mu$ represents off-diagonal (valence) gluons
which are orthogonal to $\hatm_i$, $(i=3,8)$.
One can define a projectional operator which projects any color vector
$\hat V^a$ onto the Cartan plane formed by Killing vectors $\hat m_i$
\bea
&& P^{ab}= \delta^{ab}-f^{cad}f^{cbe} \hatm_i^d \hatm_i^e, \nn \\
&& P^{ab} \hat V^b = \hat m_i^a (\hat m_i \cdot \hat V).
\eea
Notice, the projectional operator is defined properly only if the vectors $\hat m_i$
satisfy the orthonormality condition.
Using this projectional operator one can easily verify that
the vectors $\hatm_i$ are covariant constant
\bea
&& \hat D_\mu \hatm_i \equiv (\pro_\mu + \hat A_\mu) \hatm_i=0,
\eea
i.e., $\hat m_i$ represent Killing vectors on $SU(3)$.

Let us consider the vector
magnetic field strength $\vec H_{\mu\nu}$ constructed from the magnetic
gauge potential
\bea
&& \vec H_{\mu \nu} = \pro_\mu \vec C_\nu - \pro_\nu \vec C_\mu
+ \vec C_\mu \times \vec C_\nu . \label{vecH}
\eea

Straightforward calculation shows
that vector magnetic field strength $\vec H_{\mu \nu}$
belongs to the Cartan plane
\bea
&& \vec H_{\mu \nu} =H^i_{\mu\nu} \hatm_i . \label{Habel}
\eea
One can check that two differential 2-forms
$H^i=dx^\mu \wedge dx^\nu H_{\mu\nu}^i$ are closed \cite{choprd80, choprl81}
\bea
dH^i=0. \label{closedform}
\eea
Due to Poincare lemma the closed magnetic field two-forms
$H^i$ are locally exact. So that the magnetic fields $H^i_{\mu\nu}$
 can be expressed explicitly in terms of dual Abelian magnetic
potential $\tC^i_{\mu}$
\bea
&& H^i_{\mu\nu} = \pro_\mu \tC^i_{\nu }-\pro_\nu \tC^i_{\mu }. \label{dualpot}
\eea
The definition of the  magnetic fields $H^i_{\mu\nu}$
implies the existence of the dual magnetic symmetry
$\tilde U(1) \times \tilde U'(1)$ which is
an essential ingredient in the dual Meissner mechanism of confinement
\bea
\delta_{\tilde U(1)} \tilde C^3_{\mu}&=& - i\pro_\mu \tilde \alpha, \nn \\
\delta_{\tilde U'(1)} \tilde C^8_{\mu }&=& - i\pro_\mu \tilde \alpha '. \label{dualsym}
\eea

The gauge invariant Abelian decomposition (\ref{Adec})
leads to the following split of the gauge field strength
into Abelian and off-diagonal parts
\bea
\vec{F}_{\mu\nu}&=& (F_{\mu \nu}^i+ H_{\mu \nu}^i) \hatm^i+ \D _\mu \vX_\nu \nn \\
&-&\D_\nu \vX_\mu + \vX_\mu \times \vX_\nu,\label{fmn}
\eea
where $F_{\mu \nu}^i = \pro_\mu A_\nu^i -\pro_\nu A_\mu^i$
is an Abelian field strength component, and $\vec X_\mu$
can be treated as a color source \cite{choprd80, choprl81}.

Let us consider an alternative parametrization
for the Killing vectors in terms of complex fields which
have a simple geometric meaning of complex projective coordinates
on the coset ${\cal M}^6=SU(3)/U(1)\times U(1)$.
The Cartan algebra of $SU(3)$ Lie algebra is generated
by two vectors ${\bf m_{3}} = \hatm_{3}^a t_{3} \, ,
{\bf m_{8}} = \hatm_{8}^a t_{8}$ with
$t_{3,8}$ as the generators in adjoint representation.
In the case of $SU(2)$ gauge theory it is known that corresponding Killing vector
can be expressed in terms of a complex $SU(2)$ vector in fundamental representation
\cite{baal,eic,adda} which can be treated as a projective coordinate
on $SU(2)/U(1)\simeq S^2$.
In $SU(3)$ gauge theory since the homogeneous space ${\cal M}^6$
possesses a global complex structure
one can define complex projective coordinates on it
by introducing two complex triplet fields $\Psi, \Phi$.
Let us first express the lowest weight vector
$\hatm_8^a$ in terms of the complex triplet
field $\Psi$ that parameterizes
the coset $  CP^2 \simeq SU(3)/SU(2)\times U(1)$
\bea
&& \hatm_8^a = -\dfrac{3}{2}\bar \Psi \lambda^a \Psi, \nn \\
&& \bar \Psi \Psi=1. \label{par1}
\eea
The definition for the vector $\hat m_8$ is consistent with the
normalization condition and symmetric $d-$product operation in
the Lie algebra of $SU(3)$
\bea
&& \hatm_8^2=1, ~~~~~~~~d^{abc}\hatm_8^b\hatm_8^c=-\dfrac{1}{\sqrt 3} \hatm_8^a.
\eea
To construct a second Cartan vector $\hatm_3$ orthogonal
to $\hatm_8$ it is convenient to define projectional
operators
\bea
&& P_{\parallel}^{ab}=\hatm_8^a \hatm_8^b ,\nn \\
&& P_\bot^{ab} = \delta^{ab} - \hatm_8^a \hatm_8^b.
\eea
 With this the vector $\hatm_3$ can be parameterized as follows
\bea
 \hatm_3^a &=& P_\bot^{ab} \bar \Phi \lambda^b \Phi
=\bar \Phi \lambda^a \Phi +\dfrac{1}{2} \bar \Psi \lambda^a \Psi, \label{par2}
\eea
where we have introduced a second complex triplet field $\Phi$.
The definition of Killing vectors $\hat m_{3,8}$ by Eqs. (\ref{par1}),(\ref{par2})
is invariant under the dual $\tilde U(1)
\times \tilde U'(1)$ local transformations
\bea
&& \Psi \rightarrow \exp [i \tilde \alpha(x) ] \Psi, \nn \\
&& \Phi \rightarrow \exp [i \tilde \alpha'(x)] \Phi, \label{dualsymm}
\eea
which represent explicitly the dual magnetic symmetry (\ref{dualsym}).
The dual magnetic potentials $\tilde C^i_{\mu}$ can be expressed through the
complex fields as follows
\bea
&& \tC^3_{\mu} = 2i (\bar \Phi \pro_\mu \Phi +\dfrac{1}{2} \bar \Psi
 \pro_\mu \Psi), \nn \\
&& \tC^8_{\mu}=2 i (-\dfrac{\sqrt 3}{2} \bar \Psi \pro_\mu \Psi). \label{dualC}
\eea

One can verify that $\hatm_i$ satisfy the following
relations
\bea
&& \hatm_i \hatm_j=\delta_{ij}, \nn \\
&& d^{abc} \hatm_i^b \hatm_j^c = d_{ijk} \hatm_k^a , \label{ortcond}
\eea
which imply the orthogonality condition for the complex fields
$\bar \Psi \Phi =0$.
One should notice, in general, it is not necessary to impose
the orthogonality condition for Killing vectors, so that
the complex fields $\Phi, \Psi$ can be treated as arbitrary
independent fields.
This implies an interesting interpretation
of the Killing vector fields as composite fields made of quarks,
i.e., the complex fields $\Phi, \Psi$
can be treated as a flavor $SU(2)$ quark doublet $(u,d)$.
A similar idea of composite chiral solitons in QCD
is considered in \cite{chernodub}.
The definition of the vectors $\hatm_i$ in terms of the complex fields
$\Psi, \Phi$ provides a minimal set of fields with
six independent degrees of freedom needed to parameterize
the homogeneous space ${\cal M}^6=SU(3)/U(1)\times U(1)$.
In the present paper for our purpose to study the Weyl symmetric structure
of the vacuum we will treat the Killing
vectors $\hat m_i$ as independent geometric objects following
the original works \cite{choprd80, choprl81, duan}.

\section{Weyl symmetric vacuum structure}

A standard approach to topological classification
of field configurations in terms of the gauge potential
$\vec A_\mu$ is based on the third
homotopy group $\pi_3 (SU(N))= \pi_3(SU(2))=Z$
which describes characteristic Chern classes
numerated by the topological Pontryagin number. In particular,
instanton solutions represent field configurations with the
minimal Euclidean action in each such a
topological class. Classical vacuum in a pure Yang-Mills theory
is defined by the equation $\vec F_{\mu\nu}=0$
which is satisfied by an arbitrary pure gauge potential $\vec A^{vac}_\mu$.
All non-equivalent topological vacuum gauge potentials
are classified by topological Chern-Simons number $n_{CS}$
\bea
n_{CS}&=&\dfrac{1}{16\pi^2}\int d^3 x \epsilon^{ijk} \Big [ A^a_i\pro_j A^a_k+
\dfrac{1}{3} f^{abc} A^a_i A^b_j A^c_k \Big ] \nn \\
  &\equiv& \dfrac{1}{16\pi^2}\int \omega_{CS}^{(3)},
\eea
where $\omega_{CS}^{(3)}$ is a differential Chern-Simons 3-form
which is closed on space of vacuum configurations of $\vec A_\mu^{vac}$.

The construction of Cho-Duan-Ge gauge invariant Abelian projection
provides a novel approach to classification of topological structure
of the classical vacuum.
It has been shown that the vacuum gauge potential in
$SU(2)$ QCD can be constructed in terms of a more simple
geometric object, the Killing vector \cite{choplb07}.
The construction of such $SU(2)$ vacuum can be easily generalized
to the case of $SU(3)$ gauge theory
\bea
\vec A_\mu^{vac}=-\tC_\mu^i \hatm_i+\vec C_\mu. \label{Avac}
\eea
One can check by direct calculation that the corresponding
field strength vanishes identically. The essential point of the vacuum construction
belongs to the property of the magnetic field strength $\vec H_{\mu\nu}$,
namely, to its appearance in the Abelianized form (\ref{Habel}).
Due to the relationship (\ref{Avac})
one can define a classical vacuum in terms of an arbitrary given
Killing vector field $\hat m_i$.

Let us first consider the case of $SU(2)$ gauge theory.
For $SU(2)$ group one has one Cartan algebra generator $t_3$ and
one Killing vector $\hat m$.
Non-equivalent topological classes of the Killing vector
are described by homotopy groups $\pi_k(SU(2)/U(1))=Z$, ($k=2,3$),
i.e., by one topological number. Topological classes of the
pure gauge potential $\vec A_\mu^{vac}$ correspond to
the homotopy group $\pi_3(SU(2))=Z$ which implies
the vacuum classification by Chern-Simons number.
So that, we have one to one correspondence
between topological classes of Killing vectors and pure gauge potentials $\vec A_\mu^{vac}$.
The Weyl group is represented by the permutation group $Z_2$
which consists of a unit element and reflections $\hat m \leftrightarrow -\hat m$.
The vacuum gauge potential is expressed by the same relation
as (\ref{Avac}) with only one Killing vector $\hat m$ and one dual
magnetic potential $\tilde C_\mu$.
Under the Weyl reflection the vector magnetic potential $\vec C_\mu$ is invariant,
whereas the dual magnetic potential $\tC_\mu$ and $\hat m$ change
their signs to opposite ones. This implies that the vacuum
gauge potentials constructed from $\hat m$ and $-\hat m$ are
identical. Therefore, the vacuum gauge potential in $SU(2)$ QCD
forms a singlet representation of the Weyl group.

In the case of $SU(3)$ QCD the Killing vectors $\hat m_i$
describe the homogeneous space ${\cal M}^6=SU(3)/U(1)\times U(1)$.
The topological structure of the Killing vector fields
is determined by two homotopy groups $\pi_k({\cal M}^6)$, $(k=2,3)$.
The third homotopy group $\pi_3({\cal M}^6)=Z$
is equivalent to $\pi_3 (SU(3))$, so that it does not
produce a new independent topological number in addition to the Chern-Simons number.
The second homotopy group $\pi_2({\cal M}^6)=Z\times Z$ implies that
topological non-equivalent classes of Killing vector fields
are numerated by two topological charges $g^i$
\bea
g^i=\int_{S^2} H^i,
\eea
where $H^i$ are two closed 2-forms defined by Eqs. (\ref{Habel},\ref{closedform}).
Explicit expressions for the topological charges $g^i$ in terms
of winding numbers will be given below when we
introduce explicit expressions for Killing vector fields.
Since each configuration of Killing vector field determines a
vacuum gauge potential by Eq. (\ref{Avac}) one
has an additional degeneration of the classical vacuum
defined in terms of the gauge potential.

The non-trivial topological structure of Killing vector space
implies that solutions of $SU(3)$ Yang-Mills theory
are classified, in general, by two topological numbers as well.
As a simple example, we will consider monopole solutions
in pure $SU(3)$ Yang-Mills theory and in Yang-Mills-Higgs theory.
Our approach allows to study the structure of the classical vacuum
and solutions under the action of the Weyl symmetry transformation
which plays an important role in classical and quantum theory.

To study the Weyl structure of vacuum and classical solutions
we will use an explicit construction for Killing vector fields.
Let us start from the constant vectors in $SU(3)$ color space
\bea
&& \ksi_i^a=\delta^a_i, ~~~~~(i=3,8).
\eea
To obtain a more general functional form for Killing vectors
it is convenient to apply a local $SU(3)$ gauge transformation
to the constant vectors $\ksi_i$
\bea
&& \ksi_i \rightarrow \hatm_i=U \ksi_i,
\eea
where $U$ is an arbitrary matrix group element of $SU(3)$.
Notice, the Killing vectors $\hat m_3,~ \hat m_8$ are orthogonal
to each other.

Let us consider one parameter subgroup of $SU(3)$ gauge transformations
acting on $\ksi_i$ as follows
\bea
&& \ksi_3 \rightarrow \ksi_3(\delta) = \ksi_3 \cos \delta+ \ksi_8 \sin \delta,  \nn \\
&&\ksi_8 \rightarrow \ksi_8(\delta) = \ksi_3 \sin 2 \delta + \ksi_8 \cos 2 \delta,
\label{deltatrans}
\eea
where, the transformation law for the second vector $\ksi_8$
is determined by the consistence requirement with
$d-$product, namely, for any Killing vector $\hn_3$
the second vector $\hn_8$
is defined by $d-$symmetry (\ref{ortcond}).
So that, the vectors  $\ksi_i$ transform on different
representations of the group $SO(2)$.

The most general expression for the Killing vector field
$\hat n_i(\delta)$ can be obtained
from $\hat m_i$ by applying the $\delta$-transformation (\ref{deltatrans})
\bea
&& \hn_3 (\delta)= \hatm_3 \cos \delta+ \hatm_8 \sin \delta, \nn \\
&&\hn_8 (\delta)= \hatm_3 \sin 2 \delta+ \hatm_8 \cos 2 \delta. \label{ndelta}
\eea
The Killing vectors $\hat n_3, \hat n_8$
are not orthogonal to each other in general
\bea
&& \hat n_i^2=1, \nn \\
&& \hat n_3^a \hat n_8^a=\sin (3\delta).
\eea
Notice, that one has only one independent Killing vector
$\hat n_3$ since the second vector $\hat n_8$ is defined
by $d$-symmetry transformation. The orthonormality condition $\hat n_3^2=1$
and $\delta$-symmetry imply that the Killing vector $\hat n_3$ has exactly
six independent degrees of freedom. So that, $\hat n_{3}$
alone parameterizes the whole homogeneous coset space ${\cal M}^6$
in a consistent manner.

We define a generalized monopole vector field $\vec C_\mu (\delta)$
and corresponding vector magnetic field $\vec H_{\mu\nu}(\delta)$
by the same definitions (\ref{Adec}, \ref{vecH})
One can verify that the magnetic field strength
$\vec H_{\mu\nu} (\delta)$ belongs
to the Cartan plane only under a certain condition
for the angle $\delta$
\bea
&& \vec H_{\mu\nu} (\delta_k)= H_{\mu\nu}^i (\delta_k) \hn_i (\delta_k), \nn \\
&& \sin (3 \delta_k) =0, ~~~~~~~~~~\delta_k=\dfrac{\pi k}{3},
\eea
where $k=(0,1,...~5)$.
Notice, the expressions for the magnetic potential $\vec C_\mu (\delta_k)$ and field strength
$\vec H_{\mu\nu}(\delta_k)$ are the same for different angle values $\delta_k$.
The Abelian dual magnetic potentials $\tilde C_\mu^i (\delta_k)$
are defined for the respective magnetic fields $H_{\mu\nu}^i (\delta_k)$ by
\bea
&&H_{\mu\nu}^i (\delta_k)=\pro_\mu \tC_\nu^i (\delta_k) - \pro_\nu \tC_\mu^i (\delta_k).
\eea
One has the following relationship for the dual magnetic potentials $\tC_\mu^i (\delta_k)$
for different angles $\delta_k$
\bea
&& \tC_\mu^3 (\delta_k)=\tC_{0\mu}^3 \cos \delta_k +\tC_{0\mu}^8 \sin \delta_k , \nn \\
&&\tC_\mu^8 (\delta_k)=\tC_{0\mu}^3 \sin (2 \delta_k) +\tC_{0\mu}^8 \cos (2\delta_k) , \label{tCdual}
\eea
where $\tC_{0\mu}^i$ are dual magnetic potentials given at zero angle value, $\delta=0$.
 It should be stressed, that the dual magnetic potentials in the
last equations are defined up to dual magnetic transformation
$\tilde U(1)\times \tilde U'(1)$.
So that, to find explicit expressions for $\tilde C_\mu^i$
one should start from the given expressions for the magnetic field $H_{\mu\nu}^i$.

Now we can construct a vacuum sextet realizing the representation
of the Weyl symmetry group $Z_6$
\bea
\vec A_\mu^{vac} (\delta_k)=- \tC_\mu^i (\delta_k) \hn_i (\delta_k) +\vec C_\mu.
\eea
The expressions for the vacuum potential $\vec A_\mu^{vac} (\delta_k)$
and for the vacuum equation $\vec F_{\mu\nu}=0$ are highly non-linear.
However, due to Abelian structure of the vacuum gauge potential,
one can verify that any linear combination of $\vec A_\mu^{vac} (\delta_k)$
with coefficients $c_k$ satisfying the condition $\sum_k c_k=1$
represents a vacuum as well. A Weyl symmetric vacuum is
given by the symmetric linear superposition of $\vec A_\mu^{vac} (\delta_k)$
\bea
\vec A_{vac~\mu}^{Weyl}=\dfrac{1}{6} \sum_{k=0,...,5} \vec A^{vac}_{\mu}(\delta_k). \label{fullWeylvac}
\eea
One can choose angle values $\delta_k=(0, \dfrac{2 \pi}{3}, \dfrac{4 \pi}{3})$
corresponding to $I,U,V$-vacuums for corresponding $SU(2)$ subgroups of $SU(3)$.
This allows to factorize the reflection subgroup from the full Weyl group and
define a reduced Weyl symmetric vacuum
\bea
\vec A_{vac~\mu}^{Weyl}=\dfrac{1}{3} (\vec A_{vac~\mu}^I+
\vec A_{vac~\mu}^U+\vec A_{vac~\mu}^V). \label{Weylvac}
\eea
In the next section we will consider a special parametrization for
the Killing vector fields and derive the corresponding
vacuum gauge potentials. Explicit expressions for $I,U,V$-type and Weyl
symmetric vacuums are given in Appendix.
The Weyl symmetric vacuum is invariant under the basic Weyl permutation group $Z_3$
and can be useful in search of new essentially $SU(3)$
classical solutions.

\section{Ansatz for instanton solutions}

Our approach to vacuum construction in terms
of the Killing vectors on $SU(3)$ allows to
define a more general ansatz for searching possible classical
solutions in $SU(3)$ Yang-Mills theory.
In this section we apply a spherically symmetric vacuum
ansatz with two topological numbers to study possible non-trivial
instanton solutions. Let us
consider a standard 't Hooft ansatz for $n=1$ $SU(2)$ instanton \cite{bpst}:
\bea
&&\vec A_\mu = f(\rho) U\pro_\mu U^{-1}, \nn \\
&& U=\dfrac{x_4+i \vec \sigma_i x_i}{\rho}, \label{thooft}
\eea
where $x_\mu = (\vec x, x_4)$ represent Cartesian coordinates in Euclidean four dimensional
space-time, and $\vec \sigma_i$ are Pauli matrices.
The expression for the pure gauge potential $U\pro_\mu U^{-1}$ in (\ref{thooft})
can be reproduced in our approach using the expression for the $SU(2)$ vacuum gauge
potential (\ref{Avac}) with a Killing vector defined by
\bea
\hat m_3 &=& e^{-n \phi t_3} e^{-(\pi-\theta)t_2} \ksi_3 = \nn \\
&&
 \left ( \ba{c}\sin \theta\cos (n \phi) \\
         \sin \theta \sin (n \phi) \\
           -\cos \theta \\
            \ea
           \right ), \label{thooftm3}
\eea
where $\ksi_3 =(0,0,1)$ for the case of $SU(2)$ group.

In the case of $SU(3)$ gauge theory
we define a Killing vector field $\hat m_3$
by the following gauge transformation \cite{choprl80, choplb82}
\bea
\hat m_3&=&e^{-n' \phi (-\frac{1}{2} t_3+\frac{\sqrt 3}{2}t_8)} e^{\theta t_7}
\cdot e^{-(n-\frac{1}{2}n')\phi t_3} e^{-\theta t_2} \hat \xi_3 \nn \\
&&=\left (\ba{c}
\sin \theta \cos \frac{\theta}{2} \cos (n-n')\phi\\
\sin \theta \cos \frac{\theta}{2} \sin (n-n')\phi\\
\frac{1}{4}\cos \theta (3+\cos \theta )\\
\sin \theta \sin \frac{\theta}{2} \cos n\phi\\
\sin \theta \sin \frac{\theta}{2} \sin n\phi\\
-\frac{1}{2}\sin \theta \cos \theta\cos(n'\phi)\\
-\frac{1}{2} \sin \theta \cos \theta \sin(n'\phi)\\
\frac{\sqrt 3}{4} \cos \theta (1-\cos \theta)\\
 \ea \right),
 \label{hatm3}
\eea
where $n, n'$ are winding numbers corresponding to
non-trivial mapping $\pi_1(U(1)\times U(1))$.
The parameters $n, n'$ determine the topological structure
of the gauge theory and they are related to instanton and
 monopole topological charges \cite{choprl80, choplb82}.
The vector $\hat m_8$ can be obtained from $\hat m_3$ by using the
d-product
\begin{eqnarray}
\hat{m}_8=\sqrt{3} d^{abc}\hat m_3^b \hat m_3^c=
\left(
\ba{c}
0\\
0\\
\frac{\sqrt{3}}{4}(1-\cos \theta)\\
0\\
0\\
\frac{\sqrt{3}}{2} \sin \theta \cos{n^\prime\phi}\\
\frac{\sqrt{3}}{2} \sin \theta \sin{n^\prime\phi}\\
\frac{1}{4}(1+3\cos\theta)
\ea
\right) . \label{hatm8}
\end{eqnarray}
Explicit expressions for respective
$I,U,V$ and Weyl symmetric vacuum gauge potentials
 are given in Appendix.

The $I-$vacuum is defined by a pure gauge potential
corresponding to embedding $I$-type $SU(2)$ subgroup into $SU(3)$
can be written in a Weyl symmetric form $(p=1,2,3)$
\bea
&& \vec A_{\mu vac}= -\dfrac{2}{3}\sum_{p=1,2,3} (\tilde C_{\mu p} \hat m_p+
\hat m_p \times \pro_\mu \hat m_p),  \label{Ivac}
\eea
where $\tilde C_{\mu p}$ represent Weyl symmetric combinations of fields
$\tilde C_{\mu 3,8}$
\bea
&& \tC_{\mu 1}= \dfrac {1}{2} (- \tC_{\mu 3}+\sqrt 3 \tC_{\mu 8}), \nn \\
&& \tC_{\mu 2}= \dfrac {1}{2} (- \tC_{\mu 3}-\sqrt 3 \tC_{\mu 8}), \nn \\
&& \tC_{\mu 3}= \tC_{\mu 3},
\eea
and we have similar definitions for the Weyl symmetric combinations
$\hat m_p$.

Let us now consider the Weyl symmetric structure of instanton
solutions corresponding to $I,U,V$-spin $SU(2)$ subgroups of $SU(3)$.
For $I$-spin $SU(2)$ subgroup
one can introduce an instanton ansatz with three trial functions $f_p(\rho)$
\bea
\vec A_{\mu}^{I}=&& -\dfrac{2}{3}\sum_{p=1,2,3} f_p(\rho)\nn\\
&&\cdot((q_p \pro_\mu \gamma+\tilde C_{\mu p} )\hat m_p+ \hat m_p \times \pro_\mu \hat m_p),\label{Ians}
\eea
where the number parameters $q_p$ satisfy Weyl symmetry condition $q_1+q_2+q_3=0$.
In this section we use simple notations $q_1\equiv i, q_2\equiv u , q_3\equiv v$.

The (anti-) self-duality equations are
\bea
F_{\mu\nu}^a=\pm \dfrac{1}{2} \sqrt g \epsilon_{\mu\nu\rho\sigma} g^{\rho\tau} g^{\sigma\hat \xi}
F_{\tau\hat \xi}^a.
\eea
One has three independent sets
of self-duality equations:
\bea
&& F_{\rho \theta}^a=\dfrac{2}{\rho \sin \theta} F_{\phi\gamma}^a, \label{eqI} \\
&&F_{\rho \phi}^a=- \dfrac{2}{\rho \sin \theta} (F_{\theta\gamma}^a
-\cos \theta F_{\theta\phi}^a), \label{eqII} \\
&&F_{\rho \gamma}^a=\dfrac{2}{\rho \sin \theta}
(F_{\theta\phi}^a-\cos \theta F_{\theta\gamma}^a).\label{eqIII}
\eea
 The first set of Eqs. (\ref{eqI}) is most simple for solving.
One has $F_{\rho\theta}^{3,8}=F_{\phi\gamma}^{3,8}=0$. Notice, all other components
of the field strength are non-zero.
 There are only three independent equations among the remaining ones in (\ref{eqI})
 corresponding to indices $a=1$,
 $a=4$ and $a=6$. Each equation actually implies two equations since
 it has one part which does not include the dependence on the angle $\theta$
 and another part which includes terms proportional to $\cos \theta$.
It is convenient to start with
equation (\ref{eqI}) with index $a=4$ which produces the following two equations:
\bea
&&-f_1'+5f_2'+8f_3'+\dfrac{1}{3\rho} \Big[2\Big (2(n+3n')uf_1^2 \nn \\
&&+ 2(n+3n')vf_2^2+2if_3(12n-8nf_3+3n'f_3)\nn \\
&&+f_1(-12nu+(2n-3n')(u+v)f_2-(4in+3in'\nn \\
&&-8nu+3n'u)f_3)+
f_2(-12nv-(4in+3in'\nn \\
&& -8nv+3n'v)f_3)\Big )\Big]=0 , \label{eq20}\\
&&-3(f_1'-f_2') +\dfrac{2n'}{\rho} \Big [2 u f_1^2+2 v f_2^2-(i+v)f_2f_3\nn \\
&&+2if_3^2 +f_1 (-(u+v) f_2-
               (i+u)f_3)\Big ] =0, \label{eq2d}
\eea
The Eq. (\ref{eqI}) with index $a=6$ produces the following two equations:
\bea
&&5f_1'+5f_2'+2f_3' +\dfrac{2n'}{\rho}\Big [-3 u f_1^2 \nn \\
&& +f_1 \big ( 6u-2(u-v)f_2+(i-u)f_3 \big ) \nn \\
&& +f_2 \big ( 3v(-2+f_2)+(-i+v)f_3\big )\Big]=0, \label{eq30} \\
&& 3f_1'-3f_2'+\dfrac{2 n'}{3\rho}\Big[-7uf_1^2-7vf_2^2 \nn \\
&& +f_1 \big (6u+2(u+v)f_2+5if_3-uf_3 \big )\nn\\
&&+2i(-6+f_3)f_3 +f_2(6v+5if_3-vf_3)\Big]=0 \label{eq3d}.
\eea
Summing up the Eqs. (\ref{eq2d}) and (\ref{eq3d}) one obtains
the equation
\bea
n'(uf_1+vf_2+2(u+v) f_3)(f_1+f_2+4f_3-6)=0. \nn  \\
\label{eq123}
\eea
Careful analysis shows that
the last equation has a non-trivial instanton solution only when $n'=0$.

Taking into account the equation (\ref{eqI}) with index $a=1$
and Eq. (\ref{eq30}) one can obtain the following relations
\bea
&&f_1(\rho)=f_2(\rho)\equiv f(\rho) \nn \\
&& f_3'+5f'=0,   \nn \\
&& \rho f'=2 n (u+v)(f-1)(3f-4).
\eea
The solution to the last equation is
\bea
f=\dfrac{4 a^2+\rho^{2n(u+v)}}{3 a^2+\rho^{2n(u+v)}},
\eea
where $a$ is a dimensional integration constant (instanton size).
Substitution of the solution into the remaining self-duality equations
fixes the values of the parameters $n=1,~u=v=\dfrac{1}{2}$.
Finally, the solution is given by
\bea
&& f_1=f_2\equiv f=\dfrac{4 a^2+\rho^{2}}{3 a^2+\rho^{2}}, \nn \\
&& f_3\equiv g=\dfrac{-2a^2+\rho^{2}}{3 a^2+\rho^{2}}. \label{fg}
\eea
For self-dual instantons
with winding number $n= 1, n'=0$ the solution is given
by two functions $f,g$. Up to re-scaling the parameter
$a$ this is the known 't Hooft instanton solution.

The $U,V$-type $SU(2)$ embedded instanton solutions
can be obtained in a similar manner starting from the
constant color vectors
\bea
&&\hat \xi^U=(0,0,-\frac{1}{2},0,0,0,0,\frac{ \sqrt 3}{2}), \nn \\
&&\hat \xi^V=(0,0,-\frac{1}{2},0,0,0,0,-\frac{ \sqrt 3}{2}).
\eea

To obtain the instanton solution from the Weyl symmetric vacuum we
generalize the ansatz (\ref{Ians}) \bea \vec A_{\mu
}^{Weyl}=c_1(\rho) \vec A_{\mu}^{I}+c_2(\rho) \vec A_{\mu}^{U}+
c_3(\rho)\vec A_{\mu}^{V}. \label{Winst} \eea In general $\vec
A_\mu^{I,U,V}$ depend on six functions $f^{I,U,V},g^{I,U,V}$,
different winding numbers $n^{I,U,V}$ and different parameters
$q_p^{I,U,V}$. It is surprising, this ansatz produces a unique
instanton solution which is given by the sum of $I,U,V-$instantons
with the same functions $(f,g)$ given in (\ref{fg})  and
coefficients $c_1=c_2=c_3=\frac{1}{3}$ (the parameters $q_p^{I,U,V}$
are different and determined by winding number $n$ for
$I,U,V-$instanton solutions). The winding numbers for
$I,U,V-$instantons must be the same, i.e., $n^I=n^U=n^V=+1$.

Notice, one can choose three $SU(2)$ vacuums corresponding to
various values of the angle  $\delta$. All vacuums are gauge equivalent,
and the instanton solution in the Weyl symmetric form (\ref{Winst})
is gauge equivalent to $SU(2)$ embedded instanton.
For the symmetric set of angles  $\delta_k=(0, \frac{2 \pi}{3},\frac{4 \pi}{3})$
the Weyl symmetric solution coincides exactly with $I$ type $SU(2)$ instanton.
So that, all $I,U,V$ and Weyl symmetric vacuums are the same for instanton
solutions due to the condition $n'=0$.

The $U$-type instanton solution is given by
\bea
&& f_1=f_3\equiv f,~~~~ f_2\equiv g \nn \\
&& q_1=-q_2-q_3, \nn \\
&& q_2=\dfrac{1}{2n}, ~~~~q_3=-\dfrac{1}{n}, \nn \\
&& n=\pm 1,
\eea
the $V$-type instanton solution is given by
\bea
&& f_2=f_3\equiv f,~~~~ f_1\equiv g \nn \\
&& q_1=-q_2-q_3, \nn \\
&& q_2=-n, ~~~~q_3=\dfrac{1}{2n}, \nn \\
&& n=\pm 1,
\eea
where the functions $f,g$ are given by the same functions
(\ref{fg}) as in the case of $I-$instanton.

Our conclusion is the following, spherically symmetric instanton
solutions are insensitive to the presence of two topological numbers.
Even though the Chern-Simons number for each $I,U,V$ vacuum
is expressed by the sum
\bea
n_{CS}=2 p_i n+\dfrac{q_i-3 p_i}{2} n', \label{csnumber1}
\eea
the self-dual equations admit a unique solution
with a constraint $n'=0$. Our analysis is restricted by spherically symmetric
solutions, it is still possible that there might exists non-spherically
symmetric, essentially $SU(3)$ instanton which admits a non-zero value for $n'$.

Notice, that the parametrization (\ref{hatm3}) is not unique.
For instance, one can perform a different
gauge transformation of the constant vector $\ksi_3$
\bea
\hat m_3&=&e^{- w_2 \gamma (-\frac{1}{2} t_3+\frac{\sqrt 3}{2}t_8)} e^{\theta t_7}
\cdot e^{- w_1 \phi t_3} e^{-\theta t_2} \hat \xi_3,   \label{param2}
\eea
where $w_1, w_2$ are winding numbers corresponding to
rotations by angles $\phi$ and $\gamma$.
The corresponding Chern-Simons number
is the same for $I,U,V$ type vacuum gauge potentials
\bea
n_{CS}=w_1 w_2.
\eea
The parametrization (\ref{param2}) is more suitable since it does not include
the gauge parameters $p_i, q_i$ present in (\ref{csnumber1}).
The presence of different expressions for the
Chern-Simons number in terms of topological numbers of the homotopy group
$\pi_2({\cal M}^6)$ is reflection of the fact that the
topological numbers $(n,n')$ (or $w_{1,2}$), as well as the
Chern-Simons number, are not
gauge invariant (contrary to topological Pontryagin number).
One should notice, since the third homotopy group
$\pi_3 ({\cal M}^6)$ coincides with  $\pi_3 (SU(3))$
the degenerated vacuum structure rather can not be removed through
vacuum tunneling effect. We expect that such a
degenerated vacuum structure manifests itself
due to generation of monopole condensation in QCD.

\section{Weyl representation for monopoles}

In this Section we consider the Weyl symmetric structure
of singular and finite energy monopoles in $SU(3)$ QCD.
Our consideration of finite energy monopoles coincides formally
with the BPS limit of monopole solutions in Yang-Mills-Higgs theory.
However, one should stress, that in QCD one should not introduce the
Higgs potential that provides the spontaneous symmetry breaking scale.
Our point is that the mass scale parameter in monopole solutions in QCD
must represent a free parameter. The Higgs field $\Phi^a$ is treated
as a deformation of the Killing vector within the pure QCD theory.
A possible mechanism of generation of the kinetic term for the Higgs field
in the Lagrangian can be realized due to quantum dynamics of gluons.
Such a mechanism takes place in the Faddeev-Skyrme model
\cite{faddNature,battye} where the topological
Killing vector field $\hn$ gains dynamical degrees of freedom
in the effective theory of QCD \cite{choleepak}.

\vspace{2mm}
{\bf V.1. Singular monopoles}
\vspace{2mm}

Let us first consider the Weyl symmetric structure of singular monopole
solutions. Singular monopoles can be constructed
using the Killing vectors $\hn_i(\delta)$ defined by Eqs.
(\ref{ndelta},~\ref{hatm3},~\ref{hatm8}).
Under the orthogonality condition $\hat n_{3}\cdot\hat n_{8}=\sin(3\delta_k)=0$ one can calculate
vector magnetic field
$\vH_{\mu\nu}^K$ and corresponding magnetic fields $H_{i\mu\nu}^K$,
$K=(I,U,V)$, for each angle $\delta_k=(0, \frac{2 \pi}{3},\frac{4 \pi}{3})$
respectively
\bea
\vH_{\mu\nu}^K&=&\pro_\mu \vC_\nu^K-\pro_\nu \vC_\mu^K+\vC_\mu^K \times \vC_\nu^K \nn \\
 &=& H_{3\mu\nu}^K \hat n_3 + H_{8\mu\nu}^K \hat n_8,
 \eea
Explicit expressions for the magnetic fields
of $I,U,V$-type monopoles are given by the following relationships
\bea
&& H_{3\mu\nu}^{I}=(n-\dfrac{n'}{2}-n'\cos \theta) \sin \theta (\pro_\mu \theta \pro_\nu \phi-
                    \pro_\nu \theta \pro_\mu \phi), \nn \\
     && H_{8\mu\nu}^{I} = \dfrac{\sqrt 3}{2} n'\sin \theta (\pro_\mu \theta \pro_\nu \phi-
                    \pro_\nu \theta \pro_\mu \phi) , \label{Imon} \\
&& H_{3\mu\nu}^{U}=(n'-\dfrac{n}{2}+\dfrac{n'}{2}\cos \theta) \sin \theta (\pro_\mu \theta \pro_\nu \phi-
                    \pro_\nu \theta \pro_\mu \phi), \nn \\
    && H_{8\mu\nu}^{U} = \dfrac{\sqrt 3}{2} (-n+ n'\cos \theta) (\pro_\mu \theta \pro_\nu \phi-
                    \pro_\nu \theta \pro_\mu \phi) , \label{Umon} \\
&& H_{3\mu\nu}^{V}=(-\dfrac{n+n'}{2}+\dfrac{n'}{2}\cos \theta) \sin \theta (\pro_\mu \theta \pro_\nu \phi-
                    \pro_\nu \theta \pro_\mu \phi), \nn \\
    && H_{8\mu\nu}^{V} = \dfrac{\sqrt 3}{2} (n-n'- n'\cos \theta) (\pro_\mu \theta \pro_\nu \phi-
                    \pro_\nu \theta \pro_\mu \phi). \label{Vmon}
\eea
Dual Abelian magnetic potentials $\tilde C_{i\mu}^K$
can be easily derived using (\ref{dualpot}).
All field strengths $H_{3\mu\nu}^{I,U,V}$ contain
the terms proportional to $n' \cos \theta$.
These terms prevent the fulfillment of equations of motion.
To provide the above monopole configurations to
be solutions of the equations of motion
one can use the freedom in the definition
of the restricted potential $\hat A_\mu$, (\ref{Adec}).
Namely, it is enough to define the Abelian gauge potential $A_\mu^3$
("photon") in Eq. (\ref{Adec})
in an appropriate way to cancel the terms proportional to
$n' \cos \theta$ in  $H_{3\mu\nu}^{K}$. For instance,
for $I-$spin case one can define the Abelian gauge potential as follows
\cite{choprl80,choplb82}
\bea
&&A^{3I}_\mu=-\dfrac{n'}{4} \cos (2 \theta ) \pro_\mu \phi,
\eea
and similarly for $U$- and $V$-type monopole solutions.
With this, the configurations (\ref{Imon}-\ref{Vmon})
represent exact singular monopole solutions of
$I,U,V$ type.
The corresponding monopole charges are defined
by
\bea
&& g^i_K=\int \vec H_{\mu\nu} \cdot \hat n_i^K d \vec S^{\mu\nu},
\eea
It is easy to calculate
the monopole charges for the singular monopoles:
\bea
g^3_I=4 \pi(n-\dfrac{n'}{2}),   & g^8_I=4 \pi(\dfrac{\sqrt 3}{2} n'),\nn \\
g^3_U=4 \pi(n'-\dfrac{n}{2}),    & g^8_U=4 \pi(-\dfrac{\sqrt 3}{2} n), \nn \\
g^3_V=4 \pi (-\dfrac{n+n'}{2}),  & g^8_V=4 \pi(\dfrac{\sqrt 3}{2}(n-n')) \label{monchs}
 \eea
For all magnetic charges one has the same
expression for the total magnetic charge
\bea
&& g^{tot}=\sqrt {(g^3)^2+(g^8)^2}=\sqrt{n^2-n n'+n'^2}
\eea
For unit monopole  charge one has six types of monopoles
corresponding to various combinations of $n,n'$ (three other monopole
solutions are obtained by reflections from $I,U,V$ monopoles).

It should be noticed, that there is a principal difference between structures
of the monopole solutions in $SU(2)$ and $SU(3)$ Yang-Mills theories.
The singlet representation of $SU(2)$ monopoles with vanishing
monopole charge, $g=0$, can be constructed from monopole and anti-monopole solutions.
In $SU(3)$ case the colorless state with a total vanishing monopole
charge can be constructed in a different way, from three $I,U,V$ monopoles.
From the Eq. (\ref{monchs}) it follows that the total sum of $I,U,V$ monopole charges
$g_K^i$ for such a system is zero.

\vspace{2mm}
{\bf V.2. Finite energy $SU(3)$ QCD monopoles}
\vspace{2mm}

A well known finite energy monopole is given by
the 't Hooft-Polyakov monopole solution in $SU(2)$
Yang-Mills-Higgs theory \cite{thooft, polyakov,bps1b, bps1bm,bps2}.
Simple generalizations of $SU(2)$ 't Hooft-Polyakov monopole
to the case of $SU(3)$ theory were considered in
\cite{COFN, chakra, marciano,burzlaff, kunz}.
To find finite energy monopole solutions in $SU(3)$
Yang-Mills-Higgs theory with color scalar octet
one can start with the Killing
vector $\hat m_3$ given by a simple gauge transformation
\bea
\hat m_3 &=& e^{-n \phi t_3} e^{-\theta t_2} \ksi_3. \label{kill3}
\eea
For a non-trivial embedded finite energy monopole
solution with two magnetic charges corresponding to
Cartan algebra of $SU(3)$ one should rather apply a parametrization
for $\hn_3$ similar to one given in (\ref{hatm3}) which would imply two monopole
charges due to presence of two winding numbers $(n,n')$.
Unfortunately, in that case the ansatz for the gauge potential
becomes non-spherically symmetric
\bea
 \Phi^a &=& \hat m_3 F(r,\theta)+\hat m_8 G(r,\theta), \nn \\
 \vec A_\mu &=&U_\mu(r,\theta) \hat m_3+S_\mu(r,\theta) \hat m_8+ \nn \\
 &&\hat m_3 \times \pro_\mu \hat m_3 P(r,\theta)+\hat m_8 \times \pro_\mu \hat m_8 Q(r,\theta). \label{monans}
\eea
 The origin of this lies in the non-trivial homotopy group
 $\pi_2(SU(3)/U(1)\times U(1))$. Namely, in the presence of
the winding number $n'$ the singular monopole solution includes the
term with the Abelian gauge potential
$A^3_\mu \hat m_3$. This term is incompatible with spherically symmetric ansatz
for $\Phi$ due to the equation of motion for Higgs field
\bea
(\vec D \vec D \Phi)^a=\lambda \Phi^a (\Phi^2-\eta^2).
\eea
The term $A^3_\mu \hat m_3$ implies that the l.h.s. of the equation
does not belong to the Cartan plane $(\hat m_3, \hat m_8)$, so that one has to introduce
six functions with angle dependence in the ansatz (\ref{monans}).

To study the structure of monopole solutions in $SU(3)$
theory one can start with $I$-type Killing vector $\hat m_3$ defined
by the gauge transformation (\ref{kill3}).
The simplest ansatz is the following
\bea
 \Phi^a &=& \hat m_3 F(r)+\hat m_8 G(r), \nn \\
 \vec A_\mu &=&\hat m_3 \times \pro_\mu \hat m_3 P(r)
 +\hat m_8 \times \pro_\mu \hat m_8 Q(r). \label{ans1}
\eea
One can treat the functions $F(r),~ G(r)$ as length
deformations of the Killing vectors $\hat m_i$.
Due to the chosen parametrization of $\hat m_3$ the second Killing vector
is just a constant color vector $\hat m_8^a=\delta_8^a$. So, the field
$Q(r)$ can be omitted in the last equation.
We consider a standard Yang-Mills-Higgs Lagrangian in Minkowski space-time
with a flat metric $\eta_{mn}=(-+++)$
\bea
{\cal L}^{YMH}&=&-\dfrac{1}{4}Tr \vec F_{mn} \vec F^{mn}
- \dfrac{1}{2}(D_m \Phi)^a (D^m \Phi)^a \nn\\
&& -\dfrac{\lambda}{4}(\Phi^2 -v^2).
\eea
Substituting the ansatz (\ref{ans1}) into the equations of motion
leads to a system of ordinary differential equations
\bea
r^2 P''&=&(1+P)(r^2 F^2+P(P+2)), \nn \\
r^2 F''&+&2 r F'= \nn \\
&& F\big (2P(P+2)+2-\dfrac{\lambda}{3} r^2 (F^2+G^2-v^2)\big ), \nn \\
r G''&+& 2 G' = -\dfrac{\lambda}{3} G (F^2+G^2-v^2). \label{ODE1}
\eea
Let us consider the solution structure of the equations
in the BPS limit, $\lambda=0$. In the case of non BPS limit
the solution was obtained numerically in \cite{kunz}.
Notice, even though $\lambda =0$ in the BPS limit,
one has still an effect of the Higgs potential which implies the asymptotic boundary
condition $\Phi(r=\infty)=v$. With this the solution to (\ref{ODE1})
reads
\bea
&& P(r)=-1+\dfrac{vr}{\sinh (vr)}, \nn \\
&& F(r)=\dfrac{vr \coth (vr)-1}{r}, \nn \\
&& G(r)=C_1.  \label{eq:bpst}
\eea

As we mentioned above, in QCD one should not introduce any
Higgs potential since in the confinement phase the color
symmetry is unbroken, so that one has no any pre-fixed
scale parameter like the spontaneous symmetry breaking parameter $v$.
An interesting observation is that in the absence
of Higgs potential, one has still a class of finite
energy monopole solutions parameterized by
an arbitrary mass scale parameter $\mu$
\bea
&& P(r)=-1+\dfrac{\mu r}{\sinh (\mu r)}, \nn \\
&& F(r)=\dfrac{\mu r \coth (\mu r)-1}{r}, \nn \\
&& G(r)= \mu C_I. \label{su3mon}
\eea
The solution coincides formally with (\ref{eq:bpst})
up to the replacement $v \leftrightarrow \mu$.
The introduced mass scale parameter $\mu$
represents a free parameter
which is treated as an intrinsic property
of classical monopole solutions in a pure QCD
where generation of a mass scale is a
result of dynamic symmetry breaking.
The integration constant $C_I$ describes two different classes of solutions.
The zero value of the constant, $C_I=0$, implies a
vanishing function $G(r)=0$.
Such a solution corresponds to the trivial embedding $SU(2)$
monopole. Non-zero value of $C_I$ defines $I$-type embedding
of the monopole corresponding to $I$-spin subgroup $SU(2)$.
For $U$ and $V$ type monopole solutions
the respective functions $G(r)$ become non-constant functions
\cite{kunz,shnir}.
Let us consider a more general class of
monopole solutions by performing the gauge transformation
(\ref{ndelta}) for the Killing vectors in the
Cartan plane with arbitrary angle $\delta$.
This will allow to determine the structure of gauge
equivalent classes of the solutions, in particular, the Weyl symmetric
structure of $I,U,V$ monopoles.
Since the function $Q(r)$ does not play any
role, it can be omitted without loss of generality.
Substituting the gauge transformed vectors $\hat n_{3,8}$
from (\ref{ndelta}) into the ansatz (\ref{ans1})
one can obtain the following equations of motion
\bea
&& r^2 P''(r)-(1+\cos^2\delta P(r))(P(r)(2+\cos^2 \delta P(r)) \nn \\
      &&+r^2 W^2(r))=0, \nn \\
&&r^2 W''(r)+2 r W'(r)- 2 W(r)(1+\cos^2 \delta P(r))^2 =0, \nn \\
&& r Z''(r)+2 Z'(r)=0, \label{eqdelta}
\eea
where we have redefined the variables in the following way
\bea
&& F(r)=-\dfrac{W(r) \cos (2 \delta)-2 Z(r) \sin \delta}{1-2 \cos (2 \delta)}, \nn \\
&& G(r)=-\dfrac{Z(r)-W(r)\sin \delta}{1-2 \cos (2 \delta)}.
\eea
For finite energy monopole configurations the equation for the
function $Z(r)$ has a constant solution
\bea
Z(r)=\mu C.
\eea
The remaining equations in (\ref{eqdelta}) after proper redefinitions
\bea
&& P(r)=\dfrac{1}{\cos^2\delta} B(r), \nn \\
&& W(r)= \dfrac{1}{\cos \delta} R(r)
\eea
reduce to equations for the functions  $B(r),~R(r)$ identical to
first two equations in (\ref{ODE1}).
Finally, the solution is given by
\bea
&& P(r)=\dfrac{1}{\cos \delta}\Big (\dfrac{\mu r}{\sin [\mu r]}-1\Big ), \nn \\
&& F(r)= \dfrac{\cos [2\delta] \sec \delta (\mu r \coth [\mu r] -1)-2 C r \sin \delta}
         {r (2 \cos (2 \delta)-1)}, \nn \\
&& G(r)=\dfrac{C r+\tan \delta (1-\mu r \coth [\mu r])}{r (2 \cos [2 \delta]-1)}. \label{sol5}
\eea

The Weyl representation for $I,U,V$ type monopoles can be obtained from the last
equations by choosing respective values for the angle
$\delta_k = (0;~\dfrac{2 \pi}{3}; ~\dfrac{4 \pi}{3})$.
One should notice, that all solutions for arbitrary angle $\delta$ are gauge equivalent.
So that, the energy density and gauge invariant magnetic flux do not depend
on angle $\delta$. The only dependent term on the integration constant $C$ is presented
by the gauge invariant term $\Phi^2$. Gauge dependent monopole charges corresponding
to the magnetic fields $H_{\mu\nu}^{3,8}=F_{\mu\nu}^a m_{3,8}^a$ include dependence
on $\delta_k, \mu$.
The solution (\ref{sol5}) implies that there are six critical values
$\delta_{cr} =\dfrac{\pi}{6} +\dfrac{\pi k}{3}$, ($k=0,1,2,...,5$),
at which solutions gain infinite energy. At these critical points
the Killing vectors $\hat n_{3,8}$ become (anti-) parallel to each other.
This implies that the variety of monopole solutions is divided into six
sectors which are separated by infinite energy barrier
and can not be connected by a smooth rotation in the Cartan plane.

\section{Discussion}

We have applied the formalism of gauge invariant Abelian projection
to study of the Weyl symmetric structure of classical $SU(3)$ QCD vacuum.
The topological structure of the vacuum can be described naturally
in terms of Killing vector fields, i.e., by the homotopy groups
$\pi_{2,3}(SU(3)/U(1)\times U(1))$. Whereas the third homotopy group
provides the topological number which is equivalent to the Chern-Simons number,
the second homotopy group implies that topologically non-equivalent
classical vacuums are described by two topological numbers in general.
We have shown that one can construct a classical vacuum for each given Killing
vector field. Topologically non-equivalent vacuums in terms of Killing vectors
form the Weyl sextet representation.
It is an interesting feature of $SU(3)$ gauge theory that it has
a non-trivial Weyl symmetric vacuum presented by a symmetric
sum of $I,U,V$-type vacuums. Our construction of a more general vacuum
can be useful in search of essentially $SU(3)$ instanton
and monopole solutions.

The Weyl symmetry in QCD manifests itself in the presence
of classical singular and finite energy monopole solutions which
form representations of the Weyl group as well. In particular,
the lowest dimension representations are given by singlet
and sextet representations. This indicates to possible existence of similar
structures of the vacuum and monopole condensates in QCD.
Indeed, it has been shown that the quantum one-loop effective potential
in QCD has a vacuum which is completely Weyl symmetric
\cite{flyv, mpla2006}.
Notice, that the orthogonality condition for the Killing vector fields
$\hat n_i$ provides a necessary minimal energy condition for the classical vacuum.
Similarly, the absolute minimum of energy in the quantum one-loop effective potential of QCD
is realized when two independent homogeneous background color magnetic fields
are orthogonal to each other \cite{ mpla2006}.

At two-loop level one might have six degenerated vacuums forming Weyl sextet
\cite{gies,nedelko}. At quantum level the infinite energy barrier
between different Weyl sectors of monopoles becomes finite \cite{gies,nedelko}.
We conjecture that two-loop effective potential has a
Weyl symmetric vacuum in addition to possible six Weyl degenerated
vacuums. To preserve the color symmetry there are two possibilities.
First one is that the depth of the Weyl symmetric vacuum is
larger than the depth of Weyl degenerated vacuums, so it provides
a true vacuum. In the case if the energy of the Weyl sextet vacuums
is lower than the energy of the Weyl symmetric vacuum
the color symmetry can still remain unbroken due to possible vacuum tunneling
which will remove the vacuum degeneracy. This will provide that Weyl
symmetry as a part of whole color symmetry remains unbroken
giving a possible answer to the problem of origin of color confinement in QCD.

\acknowledgments

The work is supported by KNRF (Grant 2010-002-1564 and
2012-002-134), NSFC (Grants 11035006 and 11175215), CAS (Contract
No. 2011T1J31), and by UzFFR (Grant F2-FA-F116).

\vspace{4 mm}

{\bf Appendix: Notations, useful relations}

\vspace{4 mm}

In consideration of instanton solutions in Section IV we use
four dimensional cylindrical polar coordinate system
\bea
&& x= \rho \cos \dfrac{\theta}{2} \cos \dfrac{\gamma+\phi}{2}, \nn \\
&& y= \rho \cos \dfrac{\theta}{2} \sin \dfrac{\gamma+\phi}{2}, \nn \\
&& z= \rho \sin \dfrac{\theta}{2} \cos \dfrac{\gamma-\phi}{2}, \nn \\
&& t= \rho \sin \dfrac{\theta}{2} \sin \dfrac{\gamma-\phi}{2}, \nn \\
&& 0\leq\theta\leq \pi,~~0\leq\phi\leq 2\pi,~~0\leq\gamma\leq 4\pi.
\eea
The corresponding metric $g_{\mu\nu}$ is given by
\bea
&& g_{\rho\rho}=1, \nn \\
&& g_{\theta\theta}=g_{\phi\phi}=g_{\gamma\gamma}=\dfrac{\rho^2}{4}, \nn\\
&& g_{\phi\gamma}=\dfrac{\rho^2 \cos \theta}{4}.
\eea

In the analysis of static monopole solutions we use
the standard three dimensional spherical coordinate system.

In the derivation of Eqn. (\ref{hatm3}) the following
relationship is used
\bea
e^{\alpha t^7}&=&1-M+M\cos{\alpha}+t^7\sin{\alpha}, \nn \\
M&=&\dfrac{1}{4}\left(
\ba{ccc}
1&0&-\sqrt 3\\
0&4&0       \\
-\sqrt 3&0&3\\
\ea\right).
\eea
Relations for other $SU(3)$ group elements can be found
in a similar manner.

\begin{widetext}

Explicit expressions for $I,~U,~V$ type vacuum gauge potentials
are given by the following equations
\bea
&&\vec A_{vac~\mu}^I= \nn \\
&&\left (\ba{c}
 \dfrac{1}{4} \cos \dfrac{\theta}{2} (4 \sin[(n-n')\phi] \pro_\mu\theta
                 -\cos[(n-n')\phi](n' \pro_\mu \phi+4(u+v) \pro_\mu \gamma) \sin \theta)\\
-\dfrac{1}{4} \cos \dfrac{\theta}{2} (4 \cos[(n-n')\phi] \pro_\mu\theta
                 +\sin[(n-n')\phi](n' \pro_\mu \phi+4(u+v) \pro_\mu \gamma) \sin \theta)\\
 \dfrac{1}{32} (3(9n'-8n)\pro_\mu\phi+4(u-3v)\pro_\mu \gamma-2 \cos\theta ((4 n+n') \pro_\mu\phi
              +8(2 u+v) \pro_\mu\gamma)-\cos[2\theta] (n'\pro_\mu\phi+4(u+v)\pro_\mu\gamma))\\
 \dfrac{1}{4} \sin\dfrac{\theta}{2} (4\sin[n\phi] \pro_\mu\phi- \sin \theta \cos[n\phi](n' \pro_\mu\phi
                                +4(u+v)\pro_\mu\gamma))\\
 -\dfrac{1}{4} \sin\dfrac{\theta}{2} (4\cos[n\phi] \pro_\mu\phi+\sin \theta \sin[n\phi](n' \pro_\mu\phi
                                +4(u+v)\pro_\mu\gamma))\\
\dfrac{1}{8} (8 \sin[n' \phi] \pro_\mu\theta+\cos[n'\phi](2(2n-n')\pro_\mu\phi +4(u-v)\pro_\mu\gamma
                 +\cos\theta(n' \pro_\mu\phi+4(u+v)\pro_\mu\gamma))\sin[n'\phi]\sin \theta) \\
 \dfrac{1}{8} (-8 \cos[n' \phi] \pro_\mu\theta+\cos[n'\phi](2(2n-n')\pro_\mu\phi+ 4(u-v)\pro_\mu\gamma
                 +\cos\theta(n' \pro_\mu\phi+4(u+v)\pro_\mu\gamma))\cos[n'\phi]\sin \theta)\\
\dfrac{\sqrt 3}{96}(-3(8n+11n') \pro_\mu\phi +4(5u+v) \pro_\mu\gamma+
6((4n-3n')\pro_\mu\phi-8v\pro_\mu\gamma)\cos\theta+3(n'\pro_\mu\phi+4(u+v)\pro_\mu\gamma)\cos[2\theta])\\
 \ea \right), \nn \\
 \label{AmuI}
\eea
\bea
&&\vec A_{vac~\mu}^U=
\left (\ba{c}
\cos\dfrac{\theta}{2}(\sin[(n-n')\phi] \pro_\mu\theta+v \cos[(n-n')\phi]\sin\theta\pro_\mu\gamma) \\
 \cos\dfrac{\theta}{2}(-\cos[(n-n')\phi] \pro_\mu\theta+v \sin[(n-n')\phi]\sin\theta\pro_\mu\gamma) \\
 \dfrac{1}{8}((7n'-6n) \pro_\mu\phi-(4u+v)\pro_\mu\gamma+((n'-2n)\pro_\mu\phi
                    +4(u+2v) \cos\theta \pro_\mu\gamma +v \cos[2\theta] \pro_\mu\gamma)\\
 \sin\dfrac{\theta}{2}(\sin[n\phi] \pro_\mu\theta +v \cos[n\phi]\sin\theta \pro_\mu\gamma)\\
 \sin\dfrac{\theta}{2}(-\cos[n\phi] \pro_\mu\theta +v \sin[n\phi]\sin\theta \pro_\mu\gamma) \\
 \dfrac{1}{4}(4\sin[n'\phi]\pro_\mu\theta-\cos[n'\phi]\sin\theta)((n'-2n)\pro_\mu\phi+2(2u+v)\pro_\mu\gamma
             +2v\cos\theta \pro_\mu\gamma)  \\
  \dfrac{1}{4}(-4\cos[n'\phi]\pro_\mu\theta-\cos[n'\phi]\sin\theta)((n'-2n)\pro_\mu\phi+
  2(2u+v)\pro_\mu\gamma +2v\cos\theta \pro_\mu\gamma)\\
\dfrac{\sqrt 3}{24}(-3(2n+3n')\pro_\mu\phi-(4u+5v)\pro_\mu\gamma-3(\cos\theta ((n'-2n)\pro_\mu\phi
           +4u\pro_\mu\gamma) +v \cos[2\theta]\pro_\mu\gamma) \\
 \ea \right), \nn \\
 \label{AmuU}
\eea
\bea
&&\vec A_{vac ~\mu}^V=
\left (\ba{c}
\cos\dfrac{\theta}{2}(\sin[(n-n')\phi] \pro_\mu\theta+u \cos[(n-n')\phi]\sin\theta\pro_\mu\gamma) \\
 \cos\dfrac{\theta}{2}(-\cos[(n-n')\phi] \pro_\mu\theta+u \sin[(n-n')\phi]\sin\theta\pro_\mu\gamma) \\
 \dfrac{1}{8}((7n'-6n) \pro_\mu\phi+(3u+4v)\pro_\mu\gamma+((n'-2n)\pro_\mu\phi
                    +4(u-v) \cos\theta \pro_\mu\gamma +u \cos[2\theta] \pro_\mu\gamma)\\
 \sin\dfrac{\theta}{2}(\sin[n\phi] \pro_\mu\theta +u \cos[n\phi]\sin\theta \pro_\mu\gamma)\\
 \sin\dfrac{\theta}{2}(-\cos[n\phi] \pro_\mu\theta +u \sin[n\phi]\sin\theta \pro_\mu\gamma) \\
 \dfrac{1}{4}(4\sin[n'\phi]\pro_\mu\theta+\cos[n'\phi]\sin\theta)((2n-n')\pro_\mu\phi+2(u+2v)
          \pro_\mu\gamma -2u\cos\theta \pro_\mu\gamma)  \\
 \dfrac{1}{4}(-4\cos[n'\phi]\pro_\mu\theta+\sin[n'\phi]\sin\theta)((2n-n')\pro_\mu\phi+2(u+2v)
          \pro_\mu\gamma -2u\cos\theta \pro_\mu\gamma) \\
\dfrac{\sqrt 3}{24}(-3(2n+3n')\pro_\mu\phi-(u-4v)\pro_\mu\gamma+3(\cos\theta ((2n-n')\pro_\mu\phi
           +4(u+v)\pro_\mu\gamma) -3u \cos[2\theta]\pro_\mu\gamma) \\
 \ea \right), \nn \\
 \label{AmuV}
\eea
The Weyl symmetric vacuum is written as follows
\bea
&&\vec A_{vac~\mu}^{Weyl}=\nn \\
&&
\left (\ba{c}
\dfrac{1}{12} \cos\dfrac{\theta}{2}(12 \sin[(n-n')\phi] \pro_\mu\theta-\cos[(n-n')\phi]\sin \theta
         (n' \pro_\mu\phi+4 p\pro_\mu\gamma))\\
-\dfrac{1}{12} \cos\dfrac{\theta}{2}(12 \cos[(n-n')\phi] \pro_\mu\theta+\sin[(n-n')\phi]\sin \theta
         (n' \pro_\mu\phi+4 p\pro_\mu\gamma))\\
\dfrac{1}{96}((83n'-72 n) \pro_\mu\phi-4q\pro_\mu\gamma+(6(n'-4n) \pro_\mu\phi+4(q-7p)\pro_\mu\gamma)
           \cos\theta-\cos[2\theta](n' \pro_\mu\phi+4p\pro_\mu\gamma))\\
\dfrac{1}{12}\sin\dfrac{\theta}{2}(12 \sin[n\phi] \pro_\mu\theta-\cos[n\phi]\sin\theta(n'\pro_\mu\phi
                          +4p\pro_\mu\gamma))\\
-\dfrac{1}{12}\sin\dfrac{\theta}{2}(12 \cos[n\phi] \pro_\mu\theta+\sin[n\phi]\sin\theta(n'\pro_\mu\phi
                          +4p\pro_\mu\gamma))\\
\sin[n'\phi]\pro_\mu\theta+\dfrac{1}{24}\cos[n'\phi]\sin\theta (6(2n-n')\pro_\mu\phi+2(p-q)\pro_\mu\gamma
+n'\cos \theta \pro_\mu\phi+4p\cos\theta \pro_\mu\gamma) \\
-\cos[n'\phi]\pro_\mu\theta+\dfrac{1}{24}\sin[n'\phi]\sin\theta (6(2n-n')\pro_\mu\phi+2(p-q)\pro_\mu\gamma
+n'\cos \theta \pro_\mu\phi+4p\cos\theta \pro_\mu\gamma) \\
\dfrac{\sqrt 3}{288}(-3(24 n+35n')\pro_\mu\phi +4(4p-q)\pro_\mu\gamma+6((12 n-7n')\pro_\mu\phi
          2(p+q) \cos \theta) \pro_\mu\gamma+3(n'\pro_\mu\phi+4p\pro_\mu\gamma)\cos[2\theta])\\
 \ea \right). \nn\\
 &&  \label{AmuW}
\eea

\end{widetext}


\begin{thebibliography}{99}

\bibitem{nambu} Y. Nambu, Phys. Rev. {\bf D10}, 4262 (1974).
\bibitem{mandelstam} S. Mandelstam, Phys. Rep. {\bf 23C}, 245 (1976).
\bibitem{polyakov77} A. Polyakov, Nucl. Phys. {\bf B120}, 429 (1977).
\bibitem{shuryak} T. Schafer, E. V.  Shuryak, Rev. Mod. Phys. {\bf 70}, 323 (1998).
\bibitem{bpst} A. Belavin, A. Polykov, A. Schwartz, and Y. Tyupkin,
Phys. Lett. {\bf 59B}, 85 (1975).
\bibitem{callan} C. Callan, R. Dashen, and
D. Gross, Phys. Lett. B63, 334 (1976).
\bibitem{jackiw} R. Jackiw and C.
Rebbi, Phys. Rev. Lett. 37, 172 (1976).
\bibitem{choprd80} Y. M. Cho, Phys. Rev. {\bf D21}, 1080 (1980).
\bibitem{choprl81} Y. M. Cho, Phys. Rev. Lett. {\bf 46}, 302 (1981).
\bibitem{duan} Y. S. Duan and Mo-Lin Ge, Sci. Sinica {\bf 11}, 1072 (1979).
\bibitem{periw} V. Periwal, hep-th/9808127.
\bibitem{fadd1} L. Faddeev, A. J. Niemi, Phys.Lett. {\bf B449}, 214
(1999); Phys.Lett. {\bf B464}, 90 (1999).
\bibitem{shab} S. V. Shabanov, Phys. Lett. {\bf B458}, 322 (1999).
\bibitem{zhang} Y.-S. Duan, P. M. Zhang, Mod. Phys. Lett.
{\bf A17}, 2283 (2002).
\bibitem{baal} P. van Baal and A. Wipf, Phys. Lett. B515, 181 (2001).
\bibitem{choplb07} Y. M. Cho, Phys. Lett. {\bf B644}, 208 (2007).
\bibitem{goddard} P. Goddard, J. Nuyts and D. Olive, Nucl. Phys. {\bf B125}, 1 (1977).
\bibitem{choprl80} Y. M. Cho, Phys. Rev. Lett. {\bf 44}, 1115 (1980).
\bibitem{choplb82} Y. M. Cho, Phys. Lett. {\bf B115}, 125 (1982).
\bibitem{gies} J. Braun, H. Gies, J. M. Pawlowski, Phys.Lett. {\bf B684}, 262 (2010).
\bibitem{nedelko} B. V. Galilo, S. N. Nedelko, Phys. Part. Nucl. Lett. {\bf 8}, 67 (2011).
\bibitem{flyv} H. Flyvbjerg, Nucl. Phys. {\bf B176}, 379 (1980).
\bibitem{mpla2006} Y. M. Cho, D. G. Pak and J. H. Kim, Mod. Phys. Lett. A21, 2789 (2006).
\bibitem{prd02} Y. M. Cho and D. G. Pak, Phys. Rev. {\bf D65}, 074027 (2002).
\bibitem{jhep04} Y. M. Cho, M. L. Walker, and D. G. Pak, JHEP {\bf 05}, 073 (2004).
\bibitem{eic} H. Eichenherr, Nucl. Phys. {\bf B 146}, 215 (1978).
\bibitem{adda} A. D'Adda, M. Louscher and P. Di Vecchia, Nucl. Phys. {\bf B 146}, 63 (1978).
\bibitem{chernodub} M. N. Chernodub, S. M. Morozov, Phys. Rev. {\bf D74}, 054506 (2006).
\bibitem{faddNature} L. Faddeev and A. Niemi, Nature {\bf 387}, 58 (1997).
\bibitem{battye} R. A. Battye and P. M. Sutcliffe, Phys. Rev. Lett. {\bf 81}, 4798 (1998).
\bibitem{choleepak} Y. M. Cho, H. W. Lee, D. G. Pak, Phys.Lett. {\bf B525},347 (2002).
\bibitem{thooft} G. 't Hooft, Nucl. Phys. {\bf B79}, 276 (1974).
\bibitem{polyakov} A. M. Polyakov, JETP Lett. {\bf 20}, 194 (1974); Pisma Zh.Eksp.Teor.Fiz.
 {\bf 20}, 430 (1974).
 \bibitem{bps1b} E. B. Bogomol'nyi, Sov. J. Nucl. Phys. {\bf 24}, 449 (1976).
 \bibitem{bps1bm} E. B. Bogomol'nyi and M. S. Marinov, Sov. J. Nucl. Phys. {\bf 23}, 355 (1976).
 \bibitem{bps2} M. K. Prasad and C. M. Sommerfield, Phys. Rev. Lett. {\bf 35}, 760 (1975).
 \bibitem{COFN} E. Corrigan, D. I. Olive, D. B. Fairlie and J. Nuyts, Nucl. Phys.
 {\bf B106}, 475 (1976).
 \bibitem{chakra} A. Chakrabarti, Annales Poincare Phys.Theor. {\bf 23}, 235 (1975).
 \bibitem{marciano} W. J. Marciano and H. Pagels, Phys. Rev. {\bf D12}, 1093 (1975).
 \bibitem{burzlaff} J. Burzlaff, Phys. Rev. {\bf D23}, 1329 (1981).
\bibitem{kunz} J. Kunz and D. Masak, Phys. Lett. {\bf B196}, 513 (1987).
\bibitem{shnir} Ya. Shnir, Phys. Scr. {\bf 69}, 15 (2004).
\end{thebibliography}
\end{document}